# Frame Combination Techniques for Ultra High-Contrast Imaging

Joseph C. Carson[*a], Markus Feldt[a], Silvano Desidera[b], Maud Langlois[c], Franco Joos[d], David Mouillet[e], Jean-Luc Beuzit[e]

[a]Max-Planck-Institut für Astronomie, Königstuhl 17, 69117 Heidelberg, Germany; [b]Osservatorio Astronomico di Padova, Vicolo dell'Osservatorio 5, I-35122 Padova, Italy; [c]Laboratoire d'Astrophysique de Marseille, 2 place Le Verrier, 13248 Marseille, France; [d]Institut für Astronomie, ETH Zürich, 8093-Zürich, Switzerland; [e]Laboratoire d'Astrophysique de l'Observatoire de Grenoble, Université J. Fourier/CNRS, BP 53, 38041 Grenoble Cedex 9, France

## ABSTRACT

We summarize here an experimental frame combination pipeline we developed for ultra high-contrast imaging with systems like the upcoming VLT SPHERE instrument. The pipeline combines strategies from the Drizzle technique, the Spitzer IRACproc package, and homegrown codes, to combine image sets that may include a rotating field of view and arbitrary shifts between frames. The pipeline is meant to be robust at dealing with data that may contain non-ideal effects like sub-pixel pointing errors, missing data points, non-symmetrical noise sources, arbitrary geometric distortions, and rapidly changing point spread functions. We summarize in this document individual steps and strategies, as well as results from preliminary tests and simulations.

**Keywords:** high contrast imaging, high angular resolution, data reduction techniques.

## 1. INTRODUCTION

Ultra high-contrast adaptive optics observations, for applications like the direct-imaging of exo-solar planets, often involve non-conventional observing protocols like a rotating field of view, simultaneous imaging in multiple wavelengths and/or polarization bands, and painstaking efforts to maintain stellar point spread function (PSF) narrowness and stability. On the data reduction side, important emphasis is placed on well-characterizing the stellar PSF, in order for its features to be distinguished from those of a potential narrowly-separated planet.

We created an IDL pipeline that processes individual frames into a single final image, allowing for the most accurate identification of potentially imaged planets. Our pipeline performs sub-pixel interpolation/alignment, spatial filtering, geometric correction, and an iterative median/mean combination. The overall strategy is based on methods from the Drizzle technique[1], the Spitzer IRACproc package[2], and homegrown codes originally developed for Palomar Adaptive Optics data[3].

## 2. FRAME COMBINATION METHOD

Our frame combination pipeline, written essentially "from scratch" using IDL, implements the following procedures:

1) Use a bilinear-interpolation Drizzle method to sub-pixel interpolate and align all beginning (base-processed) images. Any pixel weight maps (provided from outside this pipeline's procedures) would undergo similar interpolation and alignment.
2) Apply a spatial filter to remove non-PSF-like frequencies.
3) Apply any relevant geometric corrections (supplied from outside this pipeline's procedures).
4) Generate a *best estimate image* by conducting a biased median-combination of the post-step-3 images. In this procedure, we select data points *n* places below the median, to account for the fact that astronomy outlier pixels tend to be high rather than low values. We also use a bad-pixel mask (supplied from outside this pipeline's procedures) to avoid known bad regions.

---

\* jcarson@mpia.de; phone +49 6221 528 400; fax +49 6221 528 246; www.mpia.de

5) Generate a best estimate *spatial derivative array* by sampling each *best estimate image* pixel and calculating the maximum difference between that pixel and its four nearest neighbor pixels. This procedure gives us an idea of the typical deviations between neighboring pixels.

6) Remove outlier pixels from post-step-3 individual images by throwing out any post-step-3 image pixel whose value differs from the corresponding *best estimate image* pixel by greater than $k$ times the best estimate *spatial derivative array* pixel value. We also remove bad pixels that may be flagged by the inputted bad-pixel mask.

7) Generate the final image by taking a weighted mean of the now-corrected post-step-3 individual images. (This assumes a pixel weight map was provided in step 1.)

In the following sub-sections, we describe the pipeline operations in greater detail.

## 2.1 A-Priori Assumptions

In deciding on the best strategy, we made the following a-priori assumptions:

1) We assume all beginning images have already undergone (outside of this pipeline) basic image processing, like flat-fielding, sky and/or dark subtraction, and some basic bad-pixel correction/removal.

2) We assume that the aforementioned flat-fielding and bad-pixel correction procedures have outputted a weight map, reflecting the relative reliability of each pixel.

3) For shifted and/or rotated data, we assume that information on PSF position and rotation are provided from outside our pipeline.

4) We assume that geometric correction information is provided from outside this pipeline.

5) We assume that there are always >5 data values for median-combination.

6) For maximizing high-contrast sensitivities, we assume that the dominant noise source derives from the parent-star PSF, as opposed to background noise, detector noise, or other sources.

## 2.2 Sub-Pixel Interpolation

Sub-pixel interpolation, a procedure where the "resolution" of an image is enhanced in the data reduction process, serves a variety of purposes in astronomy observing. In the cases where astronomy images are dithered along the sky by non-integer pixel amounts, sub-pixel interpolation is required to accurately mosaic or co-add images. Moreover, in cases where instrument pixels may be under-sampling the PSF, sub-pixel interpolations may allow one to effectively improve the pixel scale to a finer resolution. This has been successfully done[1,2] for observations with the Hubble Space Telescope and Spitzer Space Telescope, both of which operate instruments with under-sampled pixels.

For high-contrast planet imaging searches, improved resolution, via sub-pixel interpolation, may help one be able to identify a planet signal among the interfering parent-star PSF. This may be particularly important in cases where the parent star PSF is subtracted (in the data reduction process) in order to ease the identification of a possible planet. The better the stellar PSF is characterized, via sub-pixel interpolation or other methods, the easier should be the stellar PSF removal and subsequent planet identification. One commonly used practice (as performed on the VLT[4], Gemini North Telescope[5], HST[6], Spitzer Space Telescope[7], and others) is to observe a stellar system at different image-plane rotation angles. To combine these images into one frame, and achieve the maximum signal for a potential planet, one must de-rotate the images to align them with one another. Effective sub-pixel interpolation methods are therefore necessary to achieve this objective.

While sub-pixel interpolation can help improve the effective resolution, there is a danger of introducing adverse effects, like signal-to-noise degradation, or the introduction of false features. In an effort to maximum planet signal-to-noise, but avoid false constructions, we investigated numerous interpolation algorithms - including linear, cubic, cubic spline, sinc, FFT[8], Richardson-Lucy[9,10,11], and maximum entropy methods[12,13]. (See Fruchter & Hook 2002[1], and references therein, for definitions of linear, cubic, cubic spline, and sinc interpolations.)

In the end we selected the Drizzle[1] bilinear interpolation method. We decided against the nonlinear methods (like cubic, cubic spline, sinc, Richardson-Lucy, and maximum entropy) because they effectively sacrifice signal-to-noise ratio for resolution (as described in greater detail in Fruchter & Hook 2002[1]). While this may be preferable for high signal-to-noise situations, it is risky for use in high-noise areas, as in the case of faint planet signals drenched in the interfering signal of a parent star PSF.

For judging the mentioned FFT technique, we took into account simulations presented in Marois 2004[8] that show, for high-contrast (high noise) imaging applications, the FFT method's advantages over cubic, sinc, and cubic spline interpolations. We finally decided against the FFT method because we could find no instance where its efficacy was compared with more commonly used linear interpolation methods, which have been shown to be effective for high-contrast imaging from the ground and space[1,3,7].

The Drizzle strategy that we selected was originally developed for HST image-combination. It does not exchange signal-to-noise for resolution and is robust at dealing with non-ideal effects like under-sampled pixels, missing data points, arbitrary geometric distortions, and rapidly changing PSFs. In general, linear methods like Drizzle are considered very safe for avoiding introduction of false features. A primary driver for the original development of the Drizzle technique was the ability to deal with under-sampled pixels. Specialized high-contrast instruments, like the upcoming VLT SPHERE instrument[14], most likely do not suffer from poor sampling. However, in parts of an image where the stellar PSF is the steepest, even a highly sampling pixel scale may under-resolve the PSF shape. These regions are particularly critical for planet searches, where observers would like to extend the available search space to the narrowest semi-major axes. Thus, the ability of Drizzle to deal with under-sampled pixels may provide an advantage to even the most highly sampling instruments.

As shown in Figure 1, the Drizzle interpolation method that we use operates by first shrinking each original image pixel down to a user-selected size (but preserving all original signal). The grid of shrunken pixels, or "drops", can then be mapped onto a sub-sampled output image, taking into account relevant shifts and/or rotations. The drops then rain down onto the sub-sampled output grid, with each subsample pixel receiving signal according to the fractional overlap of the raining drop. The user must select an appropriate drop and sub-pixel size, taking into account the fact that, if either the drop or subpixel size is too small, the sub-sampled grid will have empty pixels scattered throughout. If there is a pixel weight map for the original image, this too can be mapped to the subpixel grid in a similar manner. For an $n \times n$ original grid mapped onto a sub-pixelated new grid of arbitrary dimensions, we may define the sub-pixel grid with the following equation.

$$V'(x',y') = \sum_{x=1}^{n} \sum_{y=1}^{n} V(x,y) \ a(x,y,x',y') \qquad (1)$$

In the above equation, $V'(x',y')$ represents the value of the new sub-sampled pixel at location $x',y'$. $V(x,y)$ is the value of the original grid pixel at position $x,y$. $a(x,y,x',y')$ represents the fractional overlap ($0 \leq a \leq 1$) between the sub-sample pixel at $x',y'$ and the "drop" centered at $x,y$. The above equation may be applied to either the image grid or the weight-map grid.

## 2.3 Spatial Filtering

Cosmic ray and bad-pixel effects may reveal themselves by occupying unique positions (in comparison with celestial point sources) in an image's Fourier plane. One may therefore help remove these unwanted signals by Fourier transforming an image into the pupil plane and then applying a transmission filter. This has been shown to be successful for high-contrast imaging applications with ground based telescopes (see Carson et al. 2005[3] for instance). For our spatial filter, we multiply our Fourier-transformed image by an exponential transmission function that preferentially selects PSF-like frequency signals. The exponential function includes boundary cutoffs, to ensure that true PSFs are not adversely affected. We are currently in the process of defining the optimal exponential function.

## 2.4 Outlier Rejection

One of the goals of this pipeline was to perform outlier rejection that is more effective than more conventional outlier rejection strategies like median filtering, sigma-clipping, and basic bad-pixel identification algorithms. The spatial filtering technique described in subsection 2.3 helps reach part of this objective. But steps 4, 5, 6 (from section 2) form the foundation of the outlier rejection strategy. This foundation is adapted largely from methods used in the Spitzer IRACproc package[2]. The steps utilize the spatial derivative of the image to deal with variations from a changing PSF shape or from small (sub-pixel) pointing errors. Correctly dealing with such errors is important, particularly in regions where the PSF is steepest, making accurate PSF characterization challenging. Furthermore, the effective PSF variations between frames may be significantly larger than one might expect from purely statistical variations. The spatial derivative method should be effective at dealing with such phenomena, as demonstrated in data reduction for Spitzer IRAC[2] and HST[15]. Note that our outlier rejection method is optimized for high-contrast observations, where the parent-star PSF is the dominant source of noise. This outlier rejection method should be modified for background-limited cases, like regions well separated from the parent-star PSF. (See Schuster et al. 2006 for more information on alternative outlier rejection algorithms.)

As described in Step 4, a *best estimate image* is generated by calculating the biased-median of the post-step-3 images (taking into account a bad-pixel mask). "Biased median" refers to taking the value $n$ below the median value, to deal with non-symmetrical noise sources like cosmic ray effects. From this *best estimate image*, *BEI*, a *spatial derivative array*, *SDA*, (see Step 5) may be calculated using the following equation.

$$SDA(x,y) = max\{ abs( BEI(x,y) – [ BEI(x-1,y), BEI(x+1,y), BEI(x,y+1), BEI(x,y-1) ] ) \} \tag{3}$$

As mentioned in Step 6, one may remove outlier pixels from the post-step-3 images by rejecting any post-step-3 image pixel that differs in value from the *best estimate image* pixel by more than $k$ times the corresponding pixel value in the *spatial derivative array*. In other words, we reject post-step-3 image pixels that meet the following condition.

$$| \textit{post-step-3-image}(x,y) – BEI(x,y) | > k \times SDA(x,y) \tag{4}$$

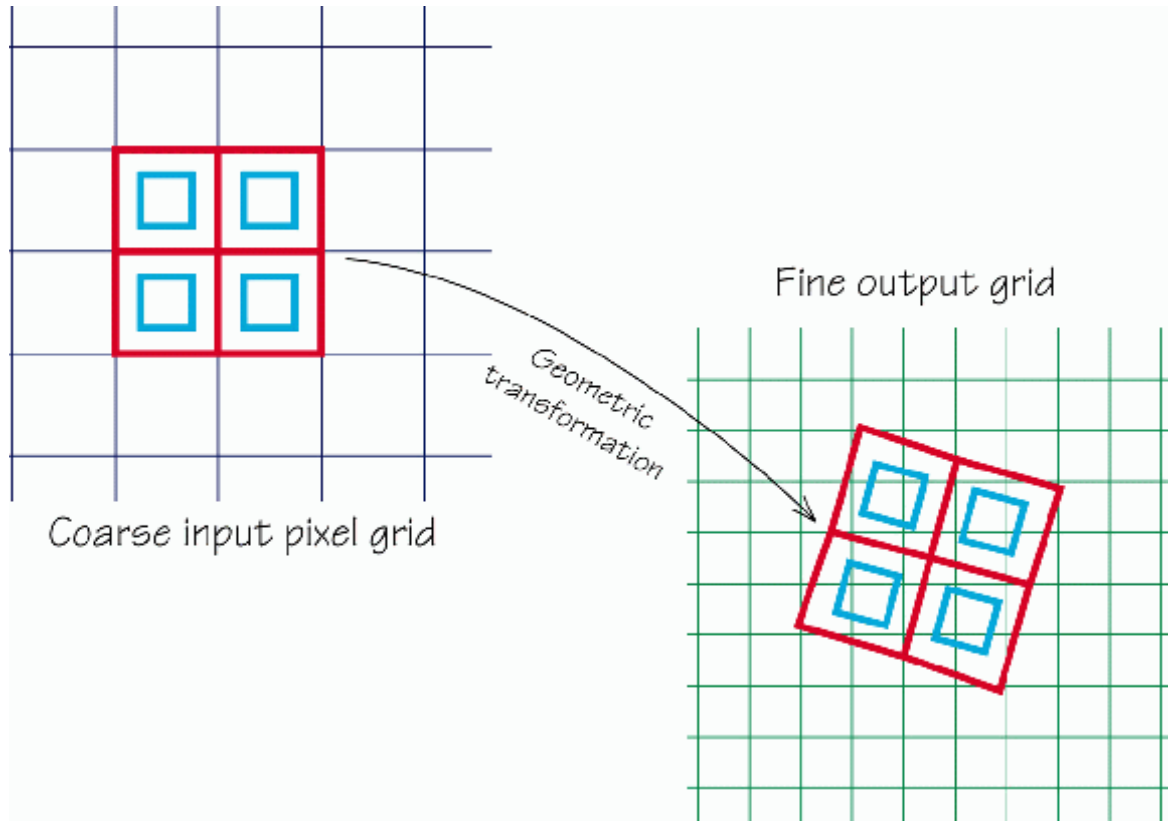

Fig. 1.** The figure above illustrates the Drizzle bilinear interpolation method that our pipeline uses. In the shown example, the signal from four original pixels (outlined in red), are concentrated into smaller square "drops" (outlined in blue). The grid of blue drops are then shifted/rotated onto an appropriate new pixel grid. The drops of signal then "rain down" onto the new grid, with each new grid pixel receiving signal according to the fractional overlap from the the raining drop.

---

\*   \*   Figure from www.stsci.edu/~fruchter/dither/drizzle.html. Reproduced here with permission from A. S. Fruchter.

## 2.5 Parameter Optimization

Our pipeline allows the user to tune various free parameters, including subsample pixel size, Drizzle "drop" size, spatial filter exponential curve, biased-median *n* parameter, and outlier rejection *k* parameter. Selecting the proper values typically involves making trade-offs between different objectives. For example, for the Drizzle operation, a smaller "drop" size helps one avoid leaving footprints of the original large pixels onto the sub-sampled array. However, if one makes the "drops" too small, the sub-pixel array will end up with missing data points. A similar conundrum exists for selecting the sub-pixel size. If one has a large stack of images to combine, with a wide spread of non-integer pixels shifts, it may be acceptable for one to aggressively sub-pixelate. However, if the number of images or spread in shifts is smaller, trying to make sub-pixels very small is inadvisable. For de-rotating images, the optimal "drop" size and sub-sample pixel size varies as function of separation from the central PSF. To achieve optimal sensitivities, one should use a different set of Drizzle parameters for exploring different regions of the image plane.

Optimal parameters for the spatial filtering algorithm depend on the particular characteristics and frequency of error sources. If such errors are effectively removed by other steps of the pipeline, or by pre-pipeline basic image processing, one might choose to not use this filter. If error sources, like high-frequency cosmic ray effects, are a significant problem, and not thoroughly addressed by other processes, one might set parameters to perform an aggressive spatial filtering. However, the user must take care not to degrade the signal-to-noise of legitimate celestial sources.

For the biased-median procedure, the user must select an optimal *n*-parameter based on how non-symmetrical are the image noise effects. For perfectly symmetrical noise and outlier pixels, *n* should be set to zero.

The optimal *k* parameter, used in the *spatial derivative array* comparison, depends on the typical magnitude and frequency of outlier pixel values. This consideration is complicated by the fact that noise sources may have different characteristics for different parts of the array – e.g. regions close to the central PSF versus regions far from the central PSF. Like for the case of the Drizzle parameters, a user might want to perform more than one reduction run, with each run using a unique parameter set to optimize a certain region of the image.

## 3. PRELIMINARY PIPELINE TESTING

To test the general effectiveness of our pipeline, we ran our code on ten simulated images that contain a stellar PSF with an implanted fake planet in the field. Our main purpose was to verify that our pipeline procedures act as expected and produce results that follow analytic predictions. In particular, we aimed to verify that the pipeline procedures in no way degrade the fundamental planet signal-to-noise level, as predicted from basic noise propagation theory. While our original simulated images include noise sources such as photon noise, sub-pixel pointing errors, and static aberrations, they do not include phenomena like cosmic ray effects and quasi-static adaptive optics speckles. For the current preliminary pipeline, we successfully verified that the results agree with analytic predictions and that there is no degradation of the implanted planet's signal-to-noise. The following sub-sections explain our tests in greater detail.

### 3.1 Initial Simulated Images

Before running our pipeline, we first generated ten simulated images with a PSF at the center. Figure 2 shows one of the ten simulated images. We aimed for our simulated images to be as similar as possible to high-contrast images expected from the upcoming SPHERE VLT instrument[14]. To create the simulated images, we first created a single pupil-plane static-aberration phase screen. We designed this phase screen to be indicative of an incoming wavefront, convolved with a model atmosphere and adaptive optics system. We Fourier transformed the final phase array to create the model imaged PSF. We next planted several fake planet PSFs, identical in shape (but not magnitude) to the central PSF. We turned this simulated image into ten images by copying the original ten times, and then adding characteristic photon noise as well as a random sub-pixel pointing error (~ 0.4 pixels). The sub-pixel shift was implemented using the Drizzle method discussed in Section 2. The final simulated images contain photon noise, but not noise deriving from a changing atmosphere. We neglected atmosphere variation effects because we believe that this phenomenon will not be a limiting factor in planet-search sensitivities. The final simulated images contain static aberrations, but not aberrations from quasi-static adaptive optics speckles. The pixel scale is 16.1 mas/pixel (~0.2 $\lambda$/D per pixel) for an assumed wavelength of 1.5 µm. We did not model the effects of cosmic ray hits.

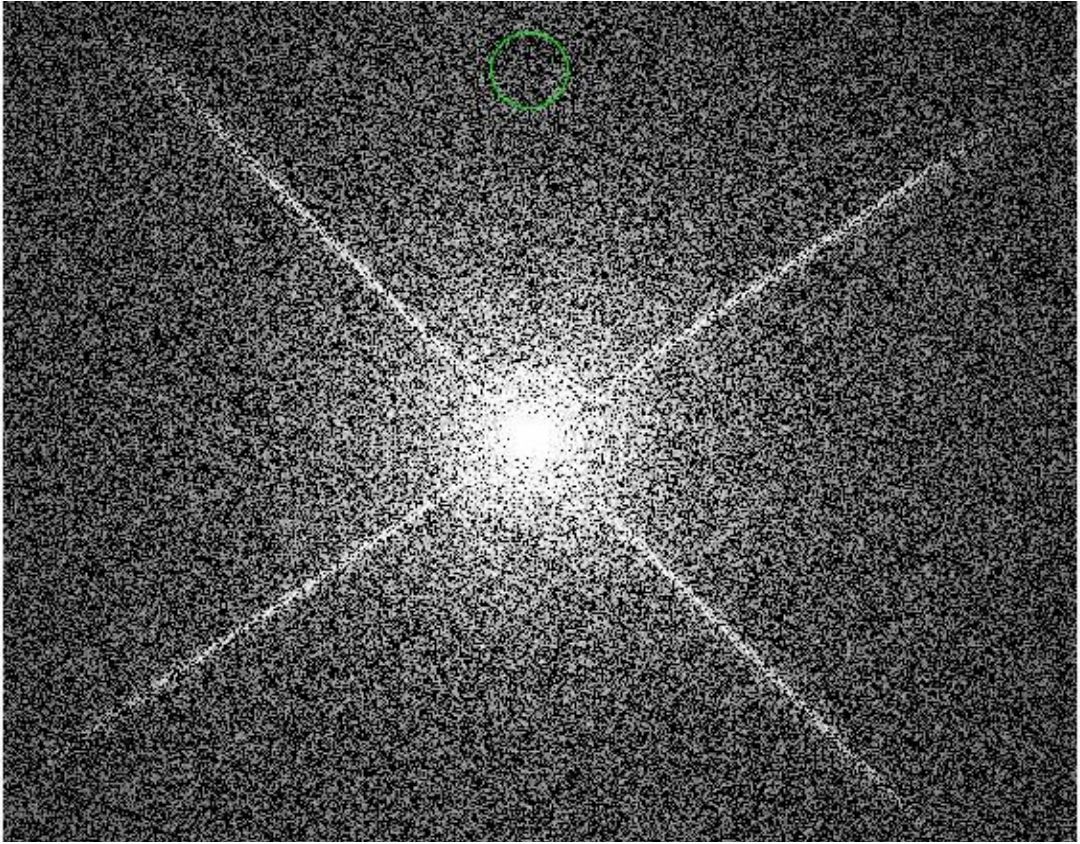

Fig. 2. One of ten simulated images that we used for testing our pipeline. The green circle (top center) surrounds an implanted simulated planet PSF (invisible amidst the noise). Modeled noise sources include photon noise, sub-pixel pointing errors and static aberrations.

### 3.2 Running the Pipeline

Inputting the ten simulated images into the pipeline, we set pipeline parameters to implement a 3-fold sub-pixel interpolation, with a "drop" width of 0.75 times the original pixel width. As our simulated images had no inter-frame rotations or dithers (other than small "unknown" pointing errors), we refrained, for now, from shifting or rotating images. Figure 3 shows a resulting sub-pixelated image.

The purpose of the spatial filter procedure is primarily to remove non-PSF-like signals like cosmic array effects and unpredictable detector pixels. Since our preliminary simulated images did not model such adverse effects, we did not implement the spatial filtering procedure. Similarly, we did not apply any geometric corrections.

Recall that the pipeline's biased median operation (step 4 in Section 2) involves the pipeline selecting a pixel value $n$ places below the median. This is useful because astronomy outliers tend to be high values rather than low values. Since our simulated images have perfectly symmetrical noise, no median bias is required; we therefore set $n = 0$. For the bad-pixel mask, we created a simulated bad-pixel map, where 0.1% of the pixels are false, and scattered randomly throughout the frame.

For pixel outlier rejection (see step 6 in Section 2), a parameter $k$ is required to set the aggressiveness of the rejection algorithm. The perfect value of $k$ depends on what part of the image one wishes to optimize – close-in to the central PSF or well-separated. In our case, we selected a baseline value of $k=5$, which seemed to result in a reasonable compromise between the two objectives. For the final weighted mean (step 7 in Section 2), we assumed an essentially flat weight map. Figure 4 shows the final image.

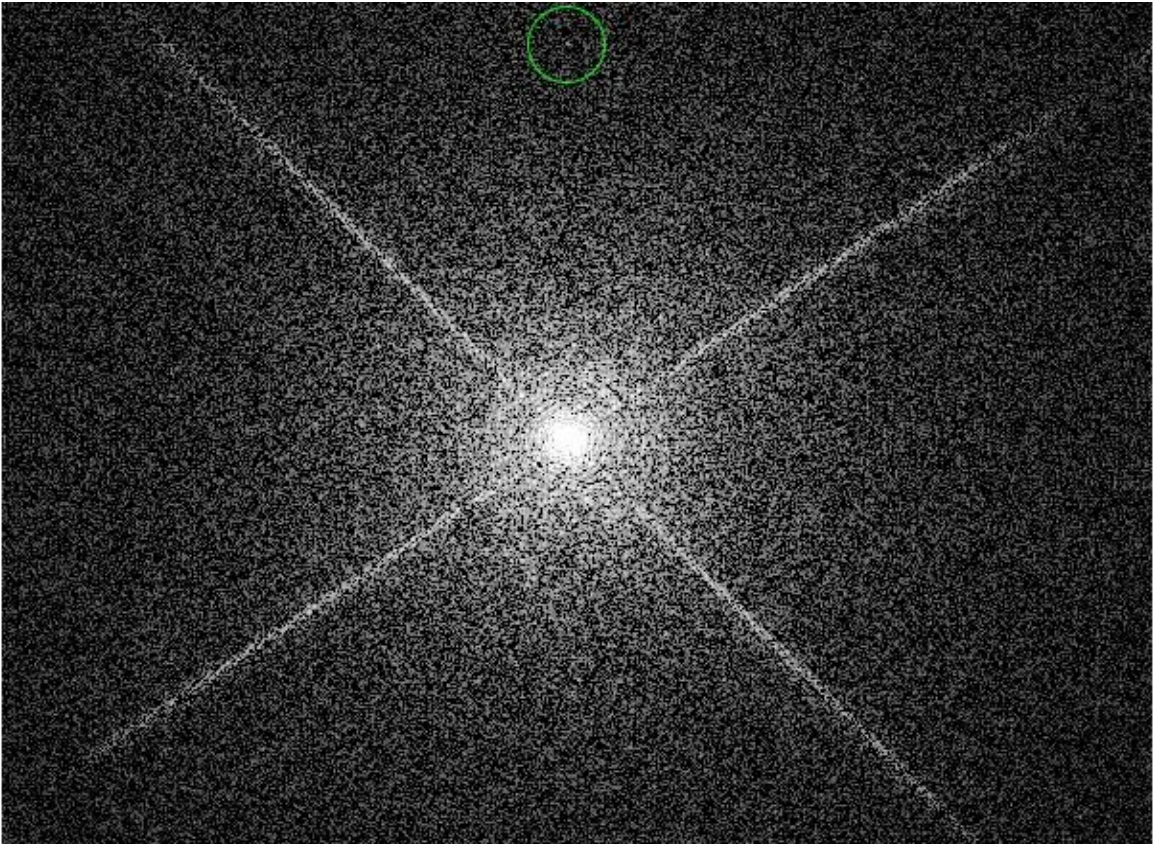

Fig. 3. The figure above represents Figure 2, after undergoing Drizzle sub-pixel interpolation (3-fold resolution increase). The green circle (top center) denotes the location of an implanted planet PSF.

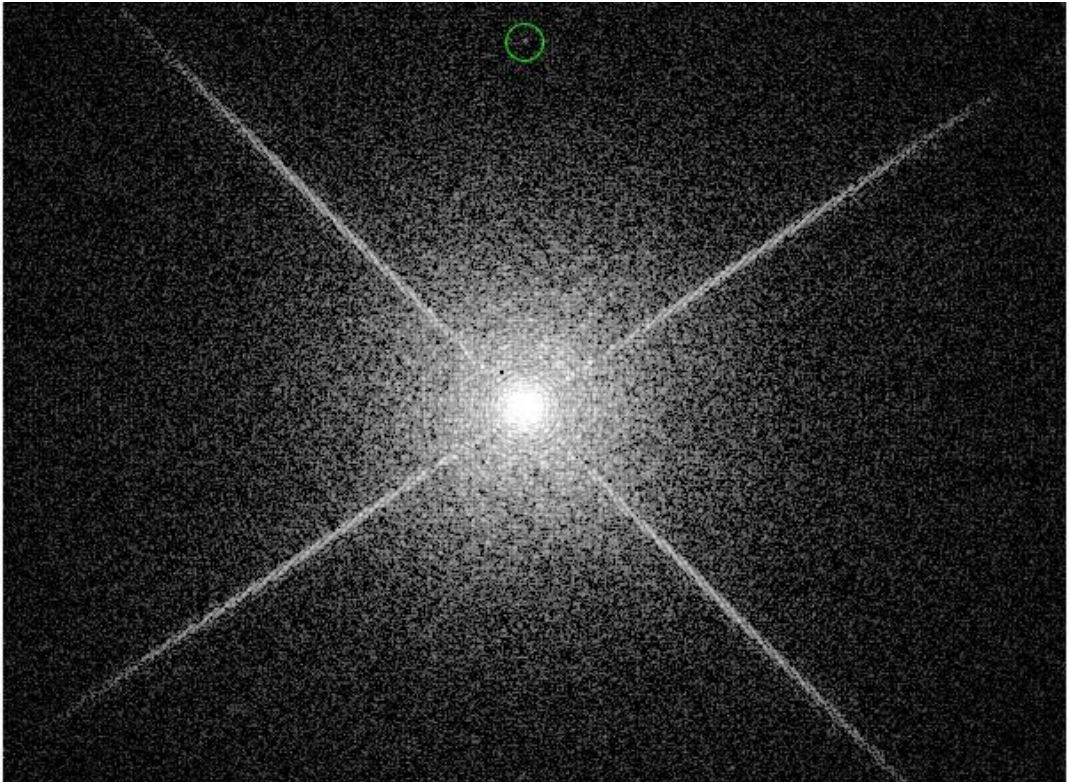

Fig. 4. Final image after all pipeline procedures have been completed. The green-circled implanted planet is now just visible (1.3-sigma level) amid the noise.

### 3.3 Pipeline Testing Results

Figure 4 shows the final image outputted by our pipeline. Using standard noise propagation theory, we verified that final noise levels are as expected. As a secondary check, we also reduced the simulated images using more conventional image combination techniques, with a basic median-combination and a standard bad-pixel removal procedure. We compared the planet signal-to-noise from the pipeline-produced image with the planet signal-to-noise from the simpler "median-combination" approach. We found that both images returned a comparable signal-to-noise level. This is re-assuring in the sense that we confirm that our pipeline causes no degradation in the planet signal-to-noise level. To understand whether the new pipeline is indeed superior (and not simply equivalent) to more basic procedures, we must wait for our implementation of quasi-static speckles and PSF subtraction procedures, cases where interpolation techniques and outlier rejections become most critical.

## 4. REMAINING WORK TO BE DONE

To thoroughly measure the efficacy of our pipeline, we would like to improve our simulated images to include additional phenomena like cosmic ray effects and quasi-static speckles. The effects of quasi-static speckles, in particular, are challenging to predict/model in simulated data. Further discussions within the SPHERE instrument group are required before we may decide on reasonable predictions. On the pipeline side, the next big step is to implement PSF subtraction capabilities. The effectiveness of PSF subtraction will give us a better measure of the effectiveness of individual steps like sub-pixel interpolation and outlier rejection. After implementing these additional features, we will likely revisit our selection of optimized parameters.

## 5. CONCLUSIONS

We developed a pipeline to perform frame combination and pixel outlier rejection for high-contrast imaging data such as those expected from the upcoming VLT SPHERE instrument. The iterative strategy was designed to be robust at dealing with data sets that may have missing data points, non-symmetrical noise sources, arbitrary geometric distortions, arbitrary inter-frame rotations/shifts, arbitrary pixel scales, and rapidly changing PSF shapes. We tested our current pipeline on simulated high-contrast data and verified that results agree with expectations from standard noise propagation theory.

## ACKNOWLEDGMENTS

We thank the IRACproc development team, including Massimo Marengo and Michael Schuster, for useful advice on high-contrast data reduction strategies.